%% file: main.tex
\documentclass[sigconf, natbib=false]{acmart}

\usepackage[utf8]{inputenc}
\usepackage{siunitx}
\usepackage{hyphenat}
\usepackage[english]{babel}

\usepackage{xcolor}
\usepackage{ifthen}

\usepackage{tikz}
\usetikzlibrary{automata, positioning, arrows, arrows.meta, decorations.pathreplacing, calc, chains, shapes.misc, backgrounds, fit, perspective, matrix, shapes, shadows.blur}

\usepackage{pgfplots}
\pgfplotsset{compat=newest}

\tikzset{
    declare function={
        normcdf(\x,\m,\s)=1/(1 + exp(-0.07056*((\x-\m)/\s)^3 - 1.5976*(\x-\m)/\s));
    }
}

\usepackage{menukeys}
\usepackage{subcaption}
\usepackage{mdframed}

\usepackage{multirow}
\usepackage{framed}

\pgfdeclarelayer{background}
\pgfsetlayers{background, main}

\usepackage{xcolor}

\definecolor{ACMYellow}{RGB}{255, 214, 0}
\definecolor{ACMOrange}{RGB}{252, 146, 0}
\definecolor{ACMRed}{RGB}{253, 27, 20}
\definecolor{ACMLightBlue}{RGB}{131, 206, 226}
\definecolor{ACMGreen}{RGB}{166, 188, 9}
\definecolor{ACMPurple}{RGB}{101, 1, 107}
\definecolor{ACMDarkBlue}{RGB}{9, 53, 122}

\definecolor{blue}{RGB}{0, 114, 178}
\definecolor{cyan}{RGB}{86, 180, 233}
\definecolor{green}{RGB}{0, 158, 115}
\definecolor{yellow}{RGB}{240, 228, 66}
\definecolor{orange}{RGB}{230,159,0}
\definecolor{red}{RGB}{213, 94, 0}
\definecolor{purple}{RGB}{204, 121, 167}

\colorlet{componentscolor}{ACMOrange}
\colorlet{participantscolor}{ACMRed}
\colorlet{teamscolor}{ACMPurple}
\colorlet{networkcolor}{ACMDarkBlue}

\usepackage[nameinlink]{cleveref}

\usepackage{enumitem}

\usepackage[style=numeric, sortcites, backend=biber]{biblatex}
\copyrightyear{2024} 
\acmYear{2024} 
\setcopyright{none} 
\acmConference[ESEM '24]{ACM/IEEE International Symposium on Empirical Software Engineering and Measurement (ESEM)}{October 20--25, 2024}{Barcelona, Spain}

\begin{document}


\title{Measuring Information Diffusion in Code Review at Spotify}

\author{Michael Dorner}
\affiliation{%
  \institution{Blekinge Institute of Technology}
  \city{Karlskrona}
  \country{Sweden}
}
\orcid{0000-0001-8879-6450}
\email{michael.dorner@bth.se}

\author{Daniel Mendez}
\orcid{0000-0003-0619-6027}
\affiliation{%
	\institution{Blekinge Institute of Technology}
	\city{Karlskrona}
	\country{Sweden}
}
\affiliation{%
	\institution{fortiss}
	\city{Munich}
	\country{Germany}
}
\email{daniel.mendez@bth.se}

\author{Ehsan Zabardast}
\affiliation{%
  \institution{Blekinge Institute of Technology}
  \city{Karlskrona}
  \country{Sweden}
}
\orcid{0000-0002-1729-5154}
\email{ehsan.zabardast@bth.se}

\author{Nicole Valdez}
\affiliation{%
  \institution{Spotify}
  \city{Stockholm}
  \country{Sweden}
}
\email{nvaldez@spotify.com}

\author{Marcin Floryan}
\affiliation{%
  \institution{Spotify}
  \city{Stockholm}
  \country{Sweden}
}
\email{mfloryan@spotify.com}

\begin{abstract}
\noindent\textbf{Background:} %
As a core practice in software engineering, the nature of code review has been frequently subject to research. Prior exploratory studies found that code review, the discussion around a code change among humans, forms a communication network that enables its participants to exchange and spread information. Although popular in software engineering, there is no confirmatory research corroborating this theory and the actual extent of information diffusion in code review is not well understood. 

\noindent\textbf{Objective:} %
In this registered report, we propose an observational study to measure information diffusion in code review to test the theory of code review as communication network. 

\noindent\textbf{Method:} %
We approximate the information diffusion in code review through the frequency and the similarity between (1) human participants, (2) affected components, and (3) involved teams of linked code reviews. The measurements approximating the information diffusion in code review serve as a foundation for falsifying the theory of code review as communication network. 



\end{abstract}

%
%

\maketitle

\section{Introduction}

The theory is compelling: Modern software systems are often too large, too complex, and evolve too fast for an individual developer to oversee all parts of the software and, thus, to understand all implications of a change. Therefore, most software projects rely on code review to foster discussions on changes and their impacts before they are merged into the code bases. During those discussions, the participants exchange information and when needed and deemed relevant, the information is passed on in subsequent code reviews. Thereby, the information diffuses in the communication network that emerges from code review. 

This theory is based on the solid and thorough exploratory research that identified information exchange as a key expectation towards code review \cite{Bacchelli2013, Baum20161, Bosu2017, Sadowski2018, Cunha2021}---also beyond teams and architectural boundaries \cite{Bacchelli2013, Baum20161, Bosu2014}---which makes code review a communication network. 

\begin{figure}
\centering
\begin{tikzpicture}
\foreach \a/\t in {90/Theory, 0/Prediction, -90/Observation, -180/Pattern}{
    \node (\t) at (\a:2cm) {\textbf{\t}};
}

\draw[-latex, thick, ACMRed] (90-25:2cm) arc (90-25:0+10:2cm);
\draw[-latex, thick, ACMRed] (0-10:2cm) arc (0-10:-90+30:2cm);
\draw[-latex, thick, ACMDarkBlue] (-90-30:2cm) arc (-90-30:-180+10:2cm);
\draw[-latex, thick, ACMDarkBlue] (-180-10:2cm) arc (-180-10:-270+25:2cm);

\draw[latex-, thick, ACMDarkBlue] (-3.5,2) -- (-3.5,-2) node[ACMDarkBlue, midway, anchor=north, rotate=90] {Inductive reasoning} node[ACMDarkBlue, midway, anchor=south, rotate=90] {Exploratory research}; 
\draw[-latex, thick, ACMRed] (3.5,2) -- (3.5,-2) node[ACMRed, midway, anchor=south, rotate=90] {Deductive reasoning} node[ACMRed, midway, anchor=north, rotate=90] {Confirmatory research}; 

\end{tikzpicture}

\caption{The empirical research cycle (in analogy to \cite{Mendez2019}): While \textcolor{ACMDarkBlue}{exploratory research} is theory-generating using inductive reasoning (starting with observations), \textcolor{ACMRed}{confirmatory research} is theory-testing using deductive reasoning (starting with a theory). This research is confirmatory.}
\label{fig:inductivedeductivecycle}
\end{figure}
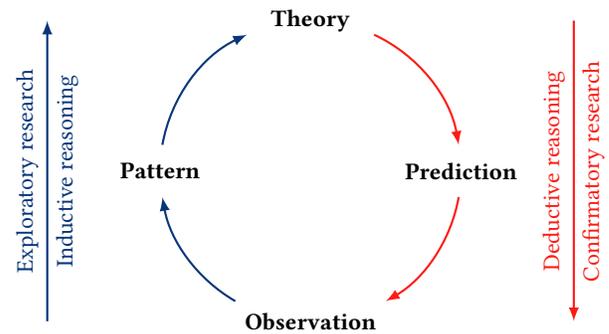

While this theory is plausible, exploratory research alone is not sufficient---it also requires the confirmatory counterpart, which is currently missing. Exploratory research begins with specific observations, distills patterns in those observations, and derives theories from the observed patterns using inductive reasoning. The nature of exploratory research leads to limited generalizability as they are drawn from specific cases. As such, it is more susceptible to researcher bias due to the absence of a predefined theory. Deductive research starts with a general theory, makes predictions (often in the form of hypotheses), and evaluates whether that prediction holds true or not in empirical observations. In research, we need both exploratory and confirmatory research to minimize bias and maximize the validity and reliability of theories efficiently. \Cref{fig:inductivedeductivecycle} shows the empirical research cycle involving both exploratory (theory-generating) and confirmatory (theory-testing) research. 

In the proposed study, we aim to fill that gap: The objective is to test the theory of code review as communication network. Instead of using classical statistical tests for the hypothesis testing, we quantify the extent of information diffusion in the code review system at Spotify, which may or may not contradict the underlying theory of code review as communication network or its universality. A single empirical code review system with no or marginal information diffusion could not be aligned with the existing theory of code review as communication network in general; further constraints, context, or limitations must be considered. Therefore, we measure information diffusion in code review at Spotify across social, organizational, and architectural boundaries.

\section{Background}
\label{sec:priorwork}

\subsection{Code Review As Communication Network}

The theory of code review as communication network is based on different exploratory studies that investigated the motivations and expectations towards code review in an industrial context \cite{Bacchelli2013, Baum20161, Bosu2017, Sadowski2018, Cunha2021}. In a synthesis of expectations and motivations towards code review reported by the exploratory studies, \citeauthor{Dorner2024} identified that information\footnote{We consistently use \emph{information} instead of knowledge as in prior work \cite{Bacchelli2013, Baum20161, Bosu2017, Sadowski2018, Cunha2021} throughout this study and concur, thereby, with \cite{Dorner2024} and \cite{Pascarella2018}. Although not equivalent, information encodes knowledge since knowledge is the meaning that may be derived from information through interpretation. This means that we may see information as a superset of knowledge. Hence, not all information is necessarily knowledge, but all knowledge is information. This allows us to subsume different stances, definitions, and notions of knowledge without an epistemological reflection upon the various definitions of knowledge. Furthermore, we can refrain from delineating the notion of knowledge from the notion of truth, the latter being too often an inherent connotation of knowledge. We may well postulate that not everything communicated is true. Opinions, expectations, misunderstandings, or best guesses are also part of any engineering and development process and do not meet knowledge and, consequently, truth by all definitions.} exchange as the cause for all effects expected from code review \cite{Dorner2024}.

\subsection{Code Review Networks}

In contrast to other work (e.g., \cite{Dorner2022, Hamasaki2013, Yang2012, Bosu2014, Thung2013}) using social network analysis to investigate code review, we use a code review network as the name suggests: a network of code reviews whose nodes represent code reviews and links indicate the references to other code reviews explicitly and manually added by human code review participants. 

This modelling approach is not novel. \citeauthor{Li2018} and \citeauthor{Hirao2019} used this modelling approach to explore the links between code reviews. \citeauthor{Hirao2019} explored the links between code reviews in six open-source projects that use Gerrit as a code review tool \cite{Hirao2019}. \citeauthor{Li2018} extended the modelling and investigation beyond pull requests to issues and identified patterns among the linkings \cite{Li2018}. Although the context (i.e., open-source software development), research objective, and analyses of those studies are not comparable, our modelling approach for the code review network, which we will discuss in \Cref{sec:measurement_model} in detail, is similar but differs as we exclude all non-human linking activities (in contrast to \cite{Li2018,Hirao2019} and use code review only (in contrast to \cite{Li2018}).

\subsection{Measuring Information Diffusion in Code Review}

Although different qualitative studies report information sharing as a key expectation towards code review \cite{Bacchelli2013, Baum20161, Bosu2017, Sadowski2018, Cunha2021}, only three prior studies have quantified information exchange in code review.

In an \emph{in-silico} experiment, \citeauthor{Dorner2024} simulated an artificial information diffusion within large (Microsoft), mid-sized (Spotify), and small code review systems (Trivago) modelled as communication networks \cite{Dorner2024}. We measured the minimal topological and temporal distances between the participants to quantify how far and how fast information can spread in code review. We found evidence that the communication network emerging from code review scales well and spreads information fast and broadly, corroborating the findings of prior qualitative work. The reported upper bound of information diffusion, however, describes information diffusion in code review under best-case assumptions, which are unlikely to be achieved. While the upper bound of information diffusion helps us already to understand the boundaries of code review as a communication network, it still does not substitute a more profound empirical measurement, for which we set the foundation with this registered report. 

In the first observational study, \citeauthor{Rigby2013} extended the expertise measure proposed by \citeauthor{Mockus2002}~\cite{Mockus2002}. The study contrasts the number of files a developer has modified with the number of files the developer knows about (submitted files $\cup$ reviewed files) and found a substantial increase in the number of files a developer knows about exclusively through code review. 

A second observational study \cite{Sadowski2018} reports (a) the number of comments per change a change author receives over tenure at Google and (b) the median number of files edited, reviewed, and both---as suggested by \citeauthor{Rigby2013} \cite{Rigby2013}. The study finds that the more senior a code change author is, the fewer code comments he or she gets. The authors ``postulate that this decrease in commenting results from reviewers needing to ask fewer questions as they build familiarity with the codebase and corroborates the hypothesis that the educational aspect of code review may pay off over time.'' In its second measurement, the study reproduces the measurements of \citeauthor{Rigby2013} but reports it over the tenure of employees at Google. They showed that reviewed and edited files are distinct sets to a large degree. 

Although the proposed file-based network creation is a sophisticated approach and may serve as a complement measurement in future studies, we found the following limitations in the measurement applied in prior work:

\begin{itemize}
\item File names may change over time, which introduces an unknown error to those measurements.
\item The software-architectural or other technical aspects (e.g., programming language, coding guidelines) of code make the measurements difficult to compare in heterogeneous software projects.  
\item We are unaware of empirical evidence that passive exposure to files in code review would lead to improved developer fluency. 
\item The explanatory power of both measurements is limited since the authors set arbitrary boundaries: \cite{Rigby2013} excluded changes and reviews that contain more than ten files, and \cite{Sadowski2018} limited the tenure of developers to 18 months and aggregated the tenure of developers by three months. 
\end{itemize}

Furthermore, our code-review-based approach differs in two aspects: First, information in code review is not only encoded in the source code but also is also in the discussions within a code review. A file-based approach does not reveal this type of information diffusion. Our code-review-based approach includes information encoded in the affected files and in the related discussions but also subsumes information on other abstraction layers of the software system. Second, a file-based approach assumes a passive and implicit information diffusion. That is, information is passively absorbed during review by the developers. In contrast, the information diffusion captured by a code-review-based approach like ours is an active information diffusion, that is, a developer actively and explicitly links information that she or he deems to be worth linking, which makes linking a human, explicit, and active decision.

\section{Research Design}
\label{sec:researchdesign}

We designed this study as an observational study~\cite{Ayala2022}  measuring the information diffusion in code review at Spotify. The measurement is not an end in itself but serves as the foundation for our hypothesis test\footnote{Hypothesis testing is often associated with statistical hypothesis testing, which is not applicable in our case, regardless of its flavor (frequentist or Bayesian). A statistical hypothesis test is built upon propositions in the context of a population using data drawn from a sample. We, however, do not sample in this study. As for any observational study, we do not aim for generalization but aim to describe and uncover associations and patterns without regard to causal relationships \cite{Ayala2022}.}: A single empirical code review system---Spotify's code review system, for example---with no or marginal information diffusion could not be aligned with the existing theory of code review as a communication network in general, and the theory as it stands would be falsified (\emph{reductio ad absurdum}). The theory must then be revised or reformulated more precisely (e.g., by adding limitations, constraints, or conditions).

In the next subsection, we state our hypotheses and discuss how we qualitatively reject our hypotheses outside classical statistical tests.  

\subsection{Hypotheses}

If code review is a communication network that enables the exchange of information (theory $T$) as identified by different exploratory studies \cite{Dorner2024}, then information substantially spread in code review 

\begin{itemize}
\item between code review participants (hypothesis $H_1$) and
\item between code software components (hypothesis $H_2$) and 
\item between teams (hypothesis $H_3$).
\end{itemize}

We can formulate this sentence as the propositional statement
\begin{equation}
T \implies \left(H_1 \land H_2 \land H_3\right).
\end{equation}

That means our theory $T$ can be falsified in its universality if one of our hypotheses cannot withstand an empirical measurement. Instead of defining arbitrary thresholds for rejecting our hypothesis, we propose a qualitative rejection criterion. This implies we will reject our hypotheses based on a comprehensive discussion of the observations of information diffusion in code review at Spotify. 

As for any observational study, the measurement model, measuring system, and actual measurement define the quality of the study. Therefore, we present our measurement model, the measuring system, and the actual measurement following the definitions in the \emph{International Vocabulary of Metrology} \cite{isovim} in the next subsections in detail.

%
%
%

\subsection{Measurement model}
\label{sec:measurement_model}

A \emph{measurement model} is the mathematical relation among all quantities known to be involved in a measurement. In this section, we describe the three approaches to quantifying information diffusion in code review, which are the foundation for the qualitative rejection of our hypotheses. 

We use a code review network to model information diffusion in code review. We define a code review network---in its verbatim meaning---as a network of code reviews whose nodes represent code reviews and whose links indicate a reference between code reviews, explicitly and manually added by code review participants. We argue that the explicit and manual referencing by code review participants is a strong indicator of actual information exchange from one code review to another. This assumption allows us to measure information diffusion without analyzing the specific information that was exchanged and its context. 

Mathematically, we model those code review networks as a directed graph $G = (C, R)$ where 
\begin{itemize}
\item $C$ is a set of vertices representing code reviews and 
\item $R$ is a set of edges which are ordered pairs of vertices representing the references between code reviews: \[R \subseteq \left\{ (a, b) \mid (a, b) \in C^2 \text{ and } a \neq b \right\}\] The direction of those edges represents the reference: The directed edge $(a, b)$ represents a code review $a$ referencing code review $b$.
\end{itemize}

\begin{figure}

\tikzstyle{node}=[draw, circle, minimum size=8mm, fill=white]

\begin{tikzpicture}

\foreach \i in {1, 4}{
	\node[node] (c\i) at (3*360/5 + 360/5 * -\i:1.25cm) {$c_\i$};
}
\node[node, draw] (c2) at (3*360/5 + 360/5 * -2:1.25cm) {$c_2$};

\foreach \i in {0, 3}{
	\node[node, draw] (c\i) at (3*360/5 + 360/5 * -\i:1.25cm) {$c_\i$};
}

\draw[-latex] (c1) to[bend right=20] (c2);
\draw[-latex] (c2) to[bend right=20] (c3);
\draw[-latex] (c2) to[bend left=20] (c0);
\draw[-latex] (c3) to[bend right=20] (c2);
\draw[-latex] (c3) to[bend left=20] (c0);
\draw[-latex] (c3) to[bend left=20] (c1);

\end{tikzpicture}
\caption{An exemplary code review network with code reviews as vertices and references between them as edges. Code reviews can reference one (see $c_1$, which references $c_2$), multiple (see $c_2$, which references $c_0$ and $c_3$), or no other code review (see $c_4$). }
\label{fig:codereviewnetwork}
\end{figure}
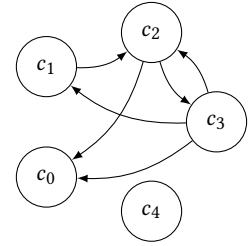

\Cref{fig:codereviewnetwork} depicts such a simple and small code review network with five code reviews linked to each other.

The relative number of linked code reviews is the first approach to quantifying information diffusion in code review and, therefore, the first input for our discussion on its significance. 

In a second approach, we approximate information diffusion in code review by measuring the similarity (or dissimilarity) of code review participants, software architecture, or organizational structure in linked code reviews: The more dissimilar the set of participants, affected code components, or involved teams of the linked code reviews, the broader the information spread in code review is.

Therefore, we enhance each code review with further information for each hypothesis: 
\begin{itemize}
\item $f_1: C \rightarrow \{\text{participants}\}$ where a code review is mapped to its participants addressing $H_1$
\item $f_2: C \rightarrow \{\text{components}\}$ where a code review is mapped to the affected components addressing $H_2$
\item $f_3: C \rightarrow \{\text{teams}\}$ where the code review is mapped to the owning teams of the affected component addressing $H_3$
\end{itemize}

Through those enhancements, we gain insights into information diffusion into three orthogonal dimensions: A social dimension, where information diffuses between code review participants; a software architectural dimension, where information diffuses software components under review; and an organizational dimension, where information diffuses between teams. Those orthogonal dimensions allow us to investigate information diffusion from different angles: Information may spread between components but may never leave the team boundaries since both components are owned by the same team. 

After enhancing, we apply two different similarity measures based on the type of enhancement to make the linked code reviews comparable along the three dimensions: 

\begin{itemize}
\item Code review participants and teams are sets. We apply the \emph{Jaccard index} to quantify the similarity between two sets. The Jaccard index (or Jaccard similarity coefficient) for two sets $A$ and $B$ is defined by 
\begin{align}
J(A, B) = \frac{|A \cap B|}{|A \cup B|} = \frac{|A \cap B|}{|A| + |B| - |A \cap B|} \;.
\end{align}

\item For the tree-like component structure, set-based operations fall short. Instead, we use the \emph{graph edit distance}, which is a measure of similarity (or dissimilarity) between two component graphs \cite{Gao2010}. The graph edit distance finds the minimal set of edit operations (insertion, deletion, substitution), in terms of cost, needed to transform one graph into another.\footnote{The graph edit distance between two graphs resembles the likely better-known string edit distance between strings: With the interpretation of strings as connected, directed acyclic graphs of maximum degree one, classical definitions of edit distance such as Levenshtein distance or Hamming distance may be interpreted as graph edit distances between suitably constrained graphs.} 

Mathematically, we define the graph edit distance as
\begin{align}
GED(G_{1},G_{2}) = \min_{(e_{1},...,e_{k}) \in \mathcal{P}(G_{1},G_{2})} \sum_{i=1}^{k} c(e_{i})
\end{align}
where $\mathcal{P}(G_{1},G_{2})$ denotes the set of edit paths transforming $G_{1}$ into (a graph isomorphic to) $G_{2}$ and $c(e) \ge 0$ is the cost of each graph edit operation $e$.
\end{itemize}

Both similarity measures are normalized, i.e., $[0, 1]$. The smaller the similarity measures, the more dissimilar the set of participants, affected code components, or involved teams of the linked code reviews. This allows us to approximate information diffusion in code review by measuring the similarity (or dissimilarity) of code review participants, software architecture, or organizational structure in linked code reviews: The smaller the similarity measures, the more dissimilar the set of participants, affected code components, or involved teams of the linked code reviews. The distribution of those similarities will indicate to what extent information spread across the boundaries mentioned before.  

In \Cref{fig:example}, we exemplify how we will use the similarities measures for discussion on falsifying the theory by three possible archetypes of cumulative distributions of all three similarity measures and their relation to the theory test. 

\begin{figure*}
\begin{subfigure}[T]{0.3\textwidth}
\centering
\begin{tikzpicture}[font=\small]
\begin{axis}[%
	xlabel=Similarity,
	width=5cm,
	grid=both,
	enlargelimits=false,
]
\addplot [smooth, participantscolor, domain=0:1, thick] {normcdf(x,0.25,0.10)};
\addplot [smooth, componentscolor, domain=0:1, thick] {normcdf(x,0.15,0.05)};
\addplot [smooth, teamscolor, domain=0:1, thick] {normcdf(x,0.22,0.08)};
\end{axis}
\end{tikzpicture}
\caption{A large dissimilarity between linked code reviews along all three dimensions implies that information diffuses beyond all three boundaries; our theory is not falsified.}
\label{fig:similar}
\end{subfigure}
\hfill
\begin{subfigure}[T]{0.3\textwidth}
\centering
\begin{tikzpicture}[font=\small]
\begin{axis}[%
	xlabel=Similarity,
	width=5cm,
	grid=both,
	enlargelimits=false,
	ymax=1,
]
\addplot [smooth, teamscolor, domain=0:1, thick] {normcdf(x,0.81,0.065)};
\addplot [smooth, participantscolor, domain=0:1, thick] {normcdf(x,0.295,0.12)};
\addplot [smooth, componentscolor, domain=0:1, thick] {normcdf(x,0.55,0.19)};
\end{axis}
\end{tikzpicture}
\caption{A large dissimilarity of \textcolor{participantscolor}{code review participants} and a large similarity of \textcolor{teamscolor}{teams} among linked code reviews imply that information diffuses within a team rather than between teams; our theory is falsified in its universality. }
\label{fig:somehowsimilar}
\end{subfigure}
\hfill
\begin{subfigure}[T]{0.3\textwidth}
\centering
\begin{tikzpicture}[font=\small]
\begin{axis}[%
	xlabel=Similarity,
	width=5cm,
	grid=both,
	enlargelimits=false,
	ymax=1,
]
\addplot [smooth, participantscolor, domain=0:1, thick] {normcdf(x,0.81,0.08)};
\addplot [smooth, componentscolor, domain=0:1, thick] {normcdf(x,0.85,0.07)};
\addplot [smooth, teamscolor, domain=0:1, thick] {normcdf(x,0.82,0.1)};
\end{axis}
\end{tikzpicture}
\caption{A large similarity among all dimensions means that information diffuses to a small extent between participants, teams, and components; our theory is falsified.}
\label{fig:notsimilar}
\end{subfigure}
\caption{Three examples of cumulative distributions of the information diffusion measured in form of similarity of linked code reviews with respect to \textcolor{participantscolor}{participants}, \textcolor{componentscolor}{components}, and \textcolor{teamscolor}{teams}: Depending on the discussions of those results, we may or may not reject our hypotheses and, thus, falsify our theory.}
\label{fig:example}
\end{figure*}

\begin{figure}
\includegraphics[trim=2cm 2cm 2cm 2cm, clip, width=\columnwidth, angle=90, origin=c]{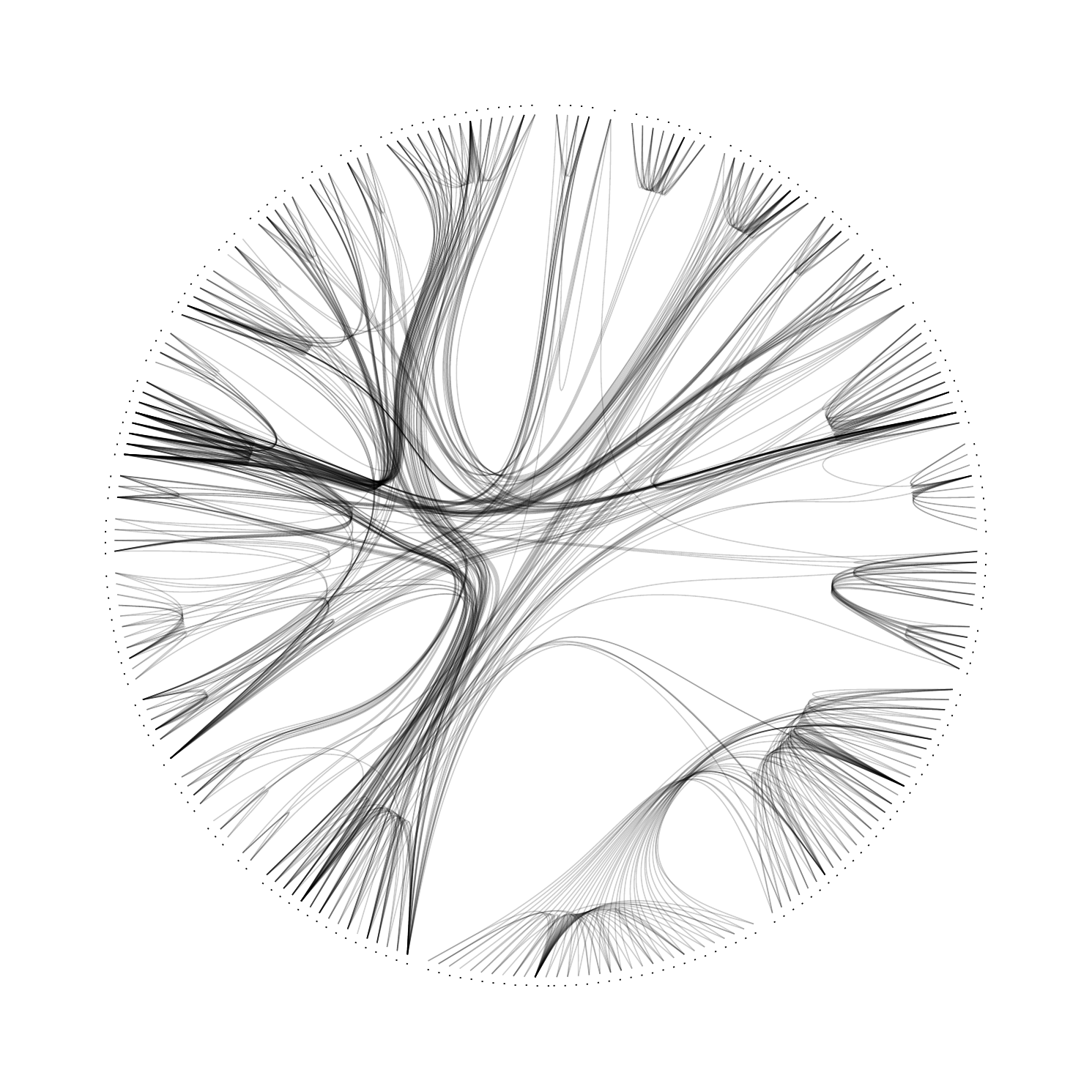}
\caption{A circular layout of the components grouped by the owning teams and linked by code reviews. This visualization may be too cluttered for massive information diffusion.}
\label{fig:circular}
\end{figure}

Aside from the two quantitative approaches, we plan to include also a visual approach. The ownership of code components allows us to cluster components per owning team, providing a more intuitive, human-comprehensive perspective. \Cref{fig:circular} uses a circular graph layout of the components grouped by the owning teams. The components are linked via the code review network $G = (C, R)$. We hope this visualization helps identify hot and cold spots and reveals the first patterns of information diffusion. 
However, depending on the extent of information between components and teams, the visualization may highlight the hot and cold spots of information diffusion, but it can also be visually overwhelming in case of a massive information diffusion.

\subsection{Measuring system}

A measuring system is the set of measuring instruments and other components assembled and adapted to give information used to generate measured values within specified intervals for quantities of specified kinds. As common in software engineering, our measuring system is a data extraction and analysis pipeline. 

\begin{figure}
\input{figures/meaurement_overview.tex}
\caption{An overview of our measuring instrument. The raw data is extracted from two different sources, Spotify's internal GitHub Enterprise (\colorbox{ACMLightBlue!20}{light blue}) and Backstage instance (\colorbox{ACMYellow!20}{yellow}). The results feed into the measurement model.}
\label{fig:measurement_system}
\end{figure}
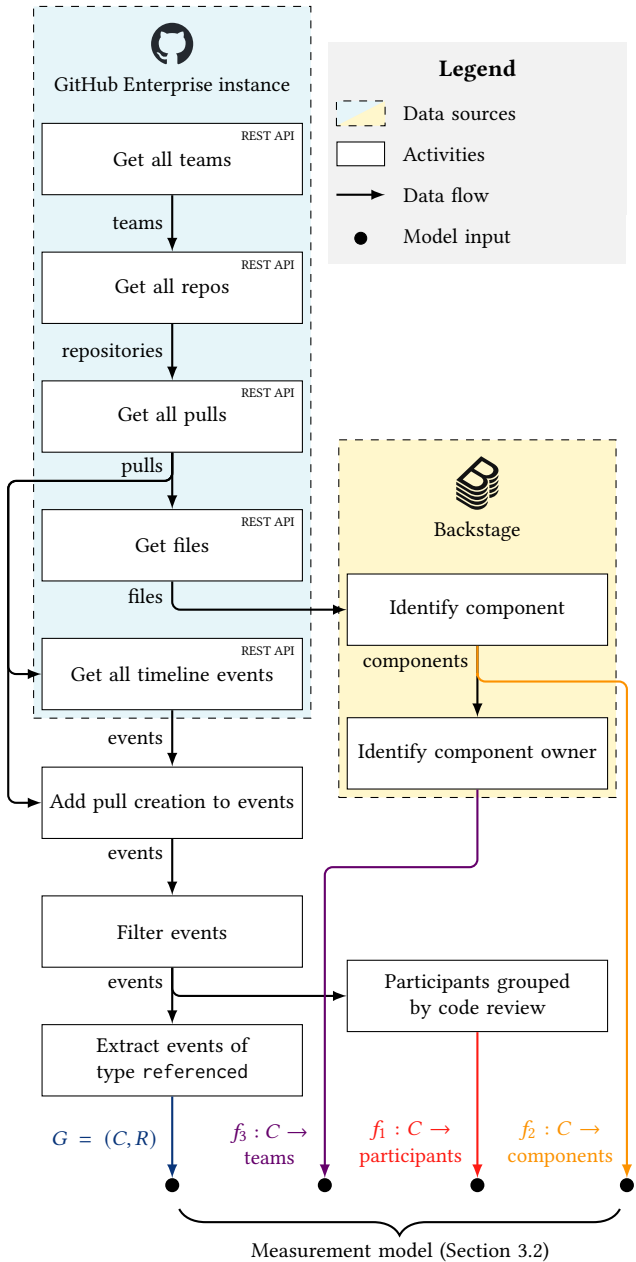

Since our measuring system is not trivial, involves a lesser-known GitHub API endpoint, and requires different data sources, we describe our measuring system in this dedicated section. \Cref{fig:measurement_system} provides a high-level overview of our measuring instrument, which we describe in detail in the following. 

The first data source for our measuring instrument is the GitHub Enterprise instance and its REST or GraphQL API. For our measurement, we follow the REST API. In GitHub, a pull request is a code review. GitHub automatically tracks\footnote{\url{https://docs.github.com/en/get-started/writing-on-github/working-with-advanced-formatting/autolinked-references-and-urls\#issues-and-pull-requests}} when a user references an issue and pull requests in such. Since internally, a code review is an issue in GitHub, we can tap the GitHub REST API endpoint for timeline events of issues\footnote{\url{https://docs.github.com/en/enterprise-server@3.10/rest/issues/timeline?apiVersion=2022-11-28\#list-timeline-events-for-an-issue}}. The timeline events contain all events triggered by activities in a pull request or issue, including the automated links to other pull requests or issue. GitHub's event endpoint \url{/events} is not suitable for extracting the event data because this API endpoint returns only a maximum of 300 events and only for the last 90 days\footnote{\url{https://docs.github.com/en/enterprise-server@3.10/rest/activity/events?apiVersion=2022-11-28\#about-github-events}}. The outcome of the crawling is a list of all events. 

Tapping the timeline events API requires the related pull requests. The GitHub search is not suitable for including or excluding pull requests since it limits its results to 1000 results per search, which is not enough at Spotify's scale. Therefore, we had to collect all pull requests from all repositories from all teams from GitHub.

We need the pull request information for two further steps: 

\begin{itemize}
\item For each pull request, we also extract all files in a pull request to map those files to components in later steps. 
\item Since there is pull request creation event available\footnote{\url{https://docs.github.com/en/enterprise-server@3.10/rest/overview/issue-event-types?apiVersion=2022-11-28}}, we add those information from the pull endpoint. 
\end{itemize}

We then filter the list of events according to the sampling frame and exclude all events from bots. 

After filtering, we extract \begin{itemize}
\item all events of type \texttt{reference}\footnote{\url{https://docs.github.com/en/enterprise-server@3.10/rest/overview/issue-event-types?apiVersion=2022-11-28\#referenced}} and its payload, the referenced pull request (code review) which results in a code review network \textcolor{ACMDarkBlue}{$G = (C, R)$}, the first input of our measurement model, and
\item all human participants grouped by each code review which results in the mapping of code review to its participants \textcolor{ACMRed}{$f_1 : C \rightarrow \text{participants}$}, a second input for our measurement model.
\end{itemize}

We believe that the GitHub referencing system is a reliable source. Two studies rely on this referencing system in GitHub \cite{Kavaler2019, Zhang2014}. However, both use the so-called \@-mentions that reference a user but not the references to issues or pull requests.


The second source for our measurement model is the software architecture description tracking all components. Spotify uses a tool called Backstage\footnote{\url{https://backstage.io}} for tracking its software architecture. For each pull request, we extracted all files and mapped the files to components. A software component is a self-contained, reusable piece of software that encapsulates the internal construction and exposes its functionality through a well-defined interface so other components can use the functionality. Software components can take many forms, including libraries, modules, classes, functions, or even entire microservices or applications. Components are hierarchically structured and may contain files or recursively other components. At Spotify, the component structure maps to the virtual folder structure of the source code. That means software components are specific folders that contain files. 

Since the component structure evolves over time, we map the files to the component structure at the time when the code reviews are referenced. Therefore, we use the available historical daily snapshots of the software architecture at Spotify. 

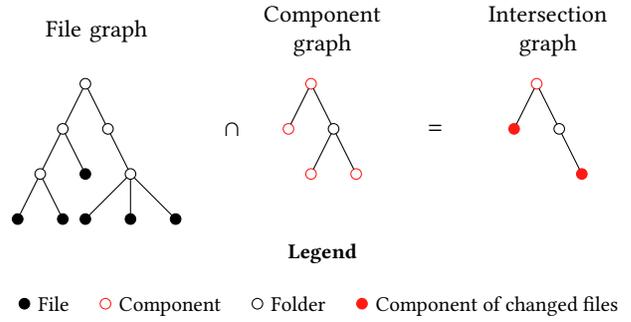
\begin{figure}
\input{figures/intersect.tex}
\caption{By intersecting the file graph (representing the changed files discussed in a code review) and the component graph (representing the component structure), we can extract the components of the changed files by extracting the leaves of the intersection graph efficiently.}
\label{fig:intersect}
\end{figure}

To identify the component of the files in a pull request efficiently, we create a file graph reflecting the paths of all changed files per code review and a time-varying component graph reflecting the component structure for each given day. The leaves of the intersection of both graphs represent the components for the files changed in a pull request. \Cref{fig:intersect} sketches the intersection of both graphs. This mapping code reviews to components \textcolor{componentscolor}{$f_2 : C \rightarrow \text{components}$} is the third input for our measurement model. 

For each identified component, we also identify its owner. Component ownership refers to the concept of assigning responsibility and accountability for a particular software component to an individual or an organizational unit within an organization. Spotify uses weak code-ownership \cite{Smite2023}. The mapping code review to owner \textcolor{teamscolor}{$f_3 : C \rightarrow \text{teams}$} is the fourth input for our measurement model.

\subsection{Measurement}

The \emph{measurement} is the process of experimentally obtaining values that can be reasonably attributed to a quantity together with any other available relevant information. 

For our measurement, we use Spotify's internal GitHub Enterprise and the Backstage instance. It comprises all Spotify-internal code reviews and components. We will run our measurement in 2024. Our sampling frame is one year and includes the timeframe $[\text{2019-01-01}, \text{2019-12-31}]$. This timeframe, outside of the ongoing developments at Spotify, allows us to publish all data in an anonymized way. However, the extent of information diffusion we will find might require us to shorten the timeframe.


%

\section{Limitations}
\label{sec:limitations}

In general, the chain of evidence of our study depends on two main factors: (1) the measurement model, measuring system, and actual measurement, and (2) the thoroughness of our discussion for qualitatively rejecting the hypotheses and, thereby, falsifying the theory of code review as communication network. 

Although we will not be able to provide the complete raw data and only a prototypical extraction pipeline for Backstage, we believe that our thorough description of our measurement model, measuring system, and the actual measurement at Spotify provides a solid foundation for this line of research. Our replication package will contain the necessary yet anonymized data to reproduce and replicate our study beyond the context of Spotify. However, as for every data-driven study, missing, incomplete, faulty, or unreliable data may significantly affect the validity of our study. To mitigate those risks, we conducted a pilot study in October 2023. Although we have not encountered such threats to validity, we cannot exclude data-related limitations. Therefore, this section will also cover the limitations that come from excluding or missing data once our data collection is completed. 

However, we believe the two most critical limitations of our study lie in the nature of a qualitative falsification of theories. Although traditional statistical hypothesis tests also have their limitations and, ultimately, also represent an implicit and qualitative discussion, we believe that a discussion remains more prone to bias, most importantly because there are no clear criteria to reject the hypotheses upfront. Such clear rejection and falsification criteria are not possible and meaningful upfront for this research; all thresholds, values, or estimates would be arbitrary. However, we believe that a comprehensive discussion makes a potential bias explicit and allows other researchers to conclude differently. Additionally, we will publish our measurement system and all intermediate anonymized data to enable other researchers to replicate our work. 

Second, even if our data and a thorough discussion suggest falsifying our theory by rejecting one of the hypotheses, our modelling approach may not capture the (relevant) information diffusion in code review. Although we have strong indications that the explicit referencing of code reviews is an active and explicit information diffusion triggered by human assessment, we are not aware of empirical evidence that supports our assumption. 

Although already discussed in \Cref{sec:researchdesign}, we emphasize again that the findings of the extent of information diffusion will not be generalizable. We do not believe that this is a major limitation of our research design since our argumentation is based on contradiction (\emph{reductio ad absurdum}).

This section will also include a detailed discussion of limitations that originate in incomplete or missing data when they become visible after the data collection and analysis.

%

\begin{acks}
We thank Spotify for supporting this research and the anonymous reviewers for their valuable and extensive feedback. This work was supported by the KKS Foundation through the SERT Project (Research Profile Grant 2018/010) at Blekinge Institute of Technology.
\end{acks}

\printbibliography

\end{document}

%% file: figures/meaurement_overview.tex
\tikzstyle{outcome}=[draw, circle, minimum width=0.5em, fill=black, inner sep=0em]
\tikzstyle{activity}=[draw, minimum width=3.25cm, minimum height=3em, anchor=west, align=center, text width=3.25cm, font=\small, fill=white]
\begin{tikzpicture}[every node/.style={font=\small}]%

\node[anchor=west, draw=none, align=center, font=\small] (github) at (0,0) {\includegraphics[width=2em]{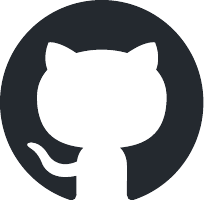}\\GitHub Enterprise instance}; 

\node[activity, below=0.25cm of github] (githubstep1) {Get all teams};

\node[activity, below=0.75 of githubstep1] (githubstep2) {Get all repos};

\node[activity, below=0.75 of githubstep2] (githubstep3) {Get all pulls};

\node[activity, below=0.75 of githubstep3]  (githubstep4) {Get files};

\node[activity, below=0.75 of githubstep4] (githubstep5) {Get all timeline events};

\node[activity, below=0.75 of githubstep5, fill=white] (addcreator) {Add pull creation to events};
\node[activity, below=0.75 of addcreator, fill=white] (filter) {Filter events};
\node[activity, below=0.75 of filter, fill=white] (createnetwork) {Extract events of type \texttt{referenced}};

\foreach \i in {1, 2, 3, 4, 5}
\node[anchor=north east, font=\tiny] at (githubstep\i.north east) {REST API};

\coordinate (c) at ($(githubstep4)!0.5!(githubstep5)$);
\coordinate (start_backstage) at (4, 0);

\node[activity, fill=white] (sscstep1) at (c-|start_backstage) {Identify component};

\node[draw=none, align=center, above=1em of sscstep1, font=\small] (ssc) 
{\includegraphics[width=2em]{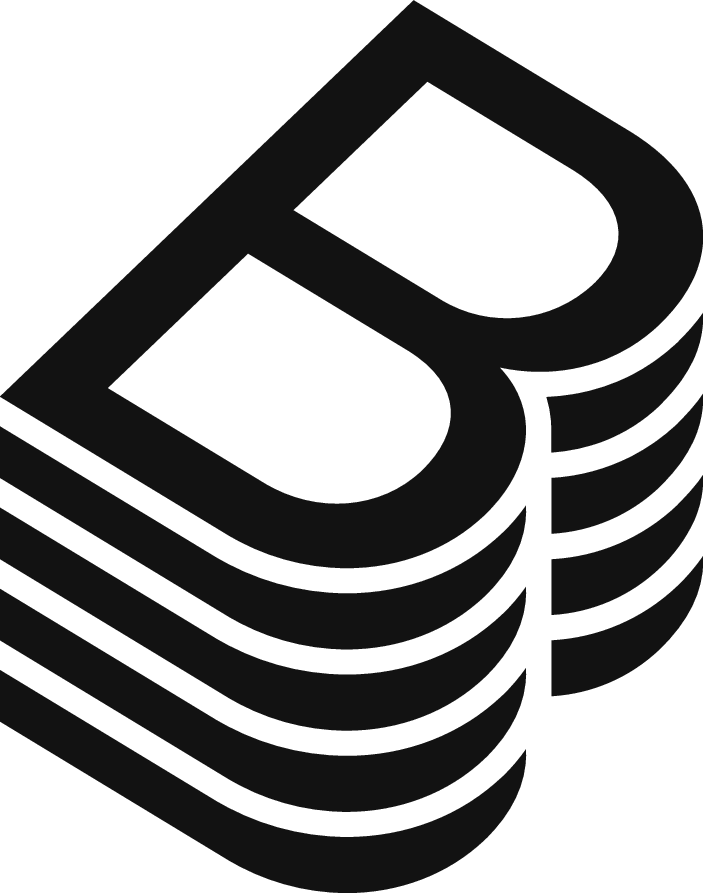}\\Backstage}; 
\node[activity, below=3em of sscstep1, fill=white] (sscstep2) {Identify component owner};

\coordinate (c) at ($(filter)!0.5!(createnetwork)$);
\node[activity, fill=white] (groupby) at (c-|sscstep2.west) {Participants grouped by code review};

\node (measurementmodel) at (0,-15) {};

\begin{scope}[on background layer]
\node[dashed, draw, fit=(ssc)(sscstep2), fill=ACMYellow!20] (githuba) {};
\end{scope}

\begin{scope}[on background layer]
\node[dashed, draw, fit=(github)(githubstep5), fill=ACMLightBlue!20] (backstagea) {};
\end{scope}


\draw[thick, -latex] (githubstep1) -- (githubstep2) node[midway, left, font=\small] {teams};
\draw[thick, -latex] (githubstep2) -- (githubstep3) node[midway, left, font=\small] {repositories};
\draw[thick, -latex] (githubstep3) -- (githubstep4) node[near start, left, font=\small] {pulls};
\draw[thick, -latex] (githubstep5.south) -- (addcreator.north) node[midway, left, font=\small] {events};
\draw[thick, -latex] (addcreator) -- (filter) node[near start, left, font=\small] {events};
\draw[thick, -latex] (filter) -- (createnetwork) node[near start, left, font=\small] {events};

\draw[thick, -latex] (sscstep1) -- (sscstep2) node[near start, left, font=\small] {components};
\draw[thick, -latex, rounded corners=3pt] (githubstep4) |- (sscstep1) node[near start, left, font=\small] {files};

\coordinate (c) at ($(githubstep3)!0.5!(githubstep4)$);

\draw[thick, -latex, rounded corners=3pt] (githubstep3) -- (c) -| (-0.5, -6) |- (githubstep5.west);

\draw[thick, -latex, rounded corners=3pt] (githubstep3) -- (c) -| (-0.5, -6) |- (addcreator.west);

\draw[thick, -latex, rounded corners=3pt] (filter) |- (groupby);

\draw[thick, -latex, ACMDarkBlue] (createnetwork) -- (createnetwork|-measurementmodel.north) node[outcome, draw=black, anchor=north, label={[font=\small, color=ACMDarkBlue, text width=1.9cm, align=right]above left:$G = (C, R)$\\$\,$}] (outcome1) {};

\coordinate (c1) at ($(addcreator)!0.5!(filter)$);
\coordinate (c2) at ($(githuba)!0.5!(backstagea)$);
\coordinate (c3) at (c1-|c2);

\draw[thick, -latex, rounded corners=3pt, ACMPurple] (sscstep2.south) -- (sscstep2|-c3) -- (c3) -- (c3|-measurementmodel.north) node[outcome, draw=black, anchor=north, label={[font=\small, color=ACMPurple, text width=1.1cm, align=center, minimum height=3em]above left:$f_3:C\rightarrow \text{teams}$}] {};

\draw[thick, -latex, rounded corners=3pt, ACMRed] (groupby.south) -- (groupby.south|-measurementmodel.north) node[outcome, draw=black, anchor=north, label={[font=\small, color=ACMRed, text width=1.4cm, align=center, minimum height=3em]above left:$f_1:C\rightarrow \text{participants}$}] {} ;

\coordinate (c) at ($(sscstep1)!0.5!(sscstep2)$);

\draw[thick, -latex, rounded corners=3pt, ACMOrange] (sscstep1) |- (c) -| ([xshift=0.25cm]sscstep2.east) -- ([xshift=0.25cm]sscstep2.east|- measurementmodel.north east) node[outcome, draw=black, anchor=north, label={[font=\small, color=ACMOrange, text width=1.4cm, align=center, minimum height=3em]above left:$f_2:C\rightarrow \text{components}$}] (outcome4) {};

\draw[thick, decorate, decoration={brace, amplitude=10pt, mirror, raise=1em}]
(outcome1) -- (outcome4) node [anchor=north, midway, yshift=-2em, font=\small, text width=4cm, align=center] {Measurement model (\Cref{sec:measurement_model})};


\matrix[anchor=north, fill=black!5, nodes={font=\small, anchor=center}, column sep={4pt}, row sep=4pt, above=13em of sscstep1,] (legend) {
\node[fill=ACMLightBlue!20, minimum width=2em, minimum height=1em] (datasources) {}; & \node[text width=2.75cm, align=left] {Data sources}; \\
\node[draw, fill=white, minimum width=2em, minimum height=1em] (activities) {}; & \node[text width=2.75cm, align=left] {Activities}; \\
\node[minimum width=2em, minimum height=1em] (dataflow) {}; & \node[text width=2.75cm, align=left] {Data flow}; \\
\node[outcome] (outcome) {}; & \node[text width=2.75cm, align=left] {Model input}; \\
};
\node[above=0em of legend, font=\bfseries, fill=black!5, minimum width=3.75cm] (legendlabel) {Legend};

\fill[ACMYellow!20] (datasources.south west) -- (datasources.north east) -- (datasources.south east) -- (datasources.south west) -- cycle;
\draw[dashed] (datasources.south west) rectangle (datasources.north east);

\draw[thick, -latex] (dataflow.west) -- (dataflow.east);

\begin{scope}[on background layer]
\node[fit=(legendlabel)(dataflow), fill=black!5] {};
\end{scope}

\end{tikzpicture}

%% file: figures/intersect.tex
\begin{tikzpicture}[
	x=1.25cm,
	folder/.style={inner sep=0mm, outer sep=0mm, circle, draw, minimum size=4pt}, 
	file/.style={folder, fill=black}, 
	component/.style={folder, ACMRed}, 
	intersection/.style={component, fill=ACMRed}, 
	legendlabel/.style={font=\small}, 
	level distance=6mm, 
	sibling distance=6mm, 
] 
\begin{scope}[local bounding box=files]
\node[folder] {}
	child { 
		node[folder] (mid) {}
		child {
			node[folder] {}
			child {
				node[file] {}
			}
			child {
				node[file] {}
			}
		}
		child {
			node[file] {} 
		}
    }
    child { 
		node[folder] {}
		child[missing] {}
		child { 
			node[folder] {} 
			child { 
				node[file] {} 
			}
			child { 
				node[file] {} 
			}			
			child { 
				node[file] {} 
			}
		}
	};
\end{scope}


\begin{scope}[local bounding box=components, xshift=3cm]
\node[component] {}
    child { 
		node[component] {} 
	}
    child { 
		node[folder] {} 
		child {
			node[component] {}
		}
		child {
			node[component] {}
		}
	};
\end{scope}

\begin{scope}[local bounding box=intersection, xshift=6cm]
\node[component] {}
    child { 
		node[intersection] (c1) {} 
	}
    child { 
		node[folder] (c2) {} 
		child[missing] { node {}}
		child { node[intersection] {}}
	};
\end{scope}

\coordinate (c1) at ($(files.east)!0.5!(components.west)$);
\coordinate (c2) at ($(components.east)!0.5!(intersection.west)$);
\node[font=\bfseries] at (mid-|c1) {$\cap$};
\node at (mid-|c2) {$=$};

\node[text width=2cm, text centered, anchor=center, yshift=2em] at (files.north) {File graph};

\node[text width=2cm, text centered, anchor=center, yshift=2em] at (components.north) {Component graph};

\node[text width=2cm, text centered, anchor=center, yshift=2em] at (intersection.north) {Intersection graph};

\end{tikzpicture}

\begin{tikzpicture}[
	folder/.style={inner sep=0mm, outer sep=0mm, circle, draw, minimum size=4pt, fill=white}, 
	file/.style={folder, fill=black}, 
	component/.style={folder, ACMRed, fill=white}, 
	intersection/.style={component, fill=ACMRed}, 
	legendlabel/.style={font=\small}, 
]

\begin{scope}[anchor=west, local bounding box=legendscope]
\matrix[row sep=0.5cm, column sep=0mm, 
    nodes={text depth=.5ex, text height=2ex},
    column 1/.style={align=left}] (m) {
   \node[file, text height=0mm] {}; & \node[legendlabel]{File}; &[3mm] \node[component, text height=0mm] {}; & \node[legendlabel]{Component}; &[3mm] 
   \node[folder, text height=0mm] {}; & \node[legendlabel]{Folder}; &&[3mm] \node[intersection, text height=0mm] {}; & \node[legendlabel]{Component of changed files}; \\
};

\node[above=1pt of m, font=\small] {\textbf{Legend}};
\end{scope}

\node[draw=none, fit=(legendscope)] {};

\end{tikzpicture}